# Chiral metamaterials with negative refractive index based on four "U" split ring resonators

Zhaofeng Li,[1] Rongkuo Zhao,[2,3] Thomas Koschny,[2,4] Maria Kafesaki,[4] Kamil Boratay Alici,[1] Evrim Colak,[1] Humeyra Caglayan,[1] Ekmel Ozbay[1,5] and C. M. Soukoulis,[2,4,a)]

[1] *Nanotechnology Research Center, and Department of Physics, Bilkent University, Bilkent, 06800 Ankara, Turkey*

[2] *Department of Physics and Astronomy and Ames Laboratory, Iowa State University, Ames, Iowa 50011, USA*

[3] *Applied Optics Beijing Area Major Laboratory, Department of Physics, Beijing Normal University, Beijing 100875, China*

[4] *Institute of Electronic Structure and Laser, Foundation for Research and Technology Hellas (FORTH), and Department of Materials Science and Technology, University of Crete, 71110 Heraklion, Greece*

[5] *Nanotechnology Research Center, Department of Physics, and Department of Electrical and Electronics Engineering, Bilkent University, Bilkent, 06800 Ankara, Turkey*

A uniaxial chiral metamaterial is constructed by double-layered four "U" split ring resonators mutually twisted by 90 degrees. It shows a giant optical activity and circular dichroism. The retrieval results reveal that a negative refractive index is realized for circularly polarized waves due to the large chirality. The experimental results are in good agreement with the numerical results.

Chiral metamaterials (CMMs) have attracted much attention ever since it was predicted that negative refraction can be achieved by a chiral route [1, 2]. In fact, chiral metamaterials also have other interesting properties, such as the giant optical activity and circular dichroism, which may find their way into optical applications [3-8]. As an artificial metamaterial, CMM lacks any mirror symmetry so that the cross-coupling between the electric and magnetic fields exists at the resonance. The degeneracy of the two circularly polarized waves is thereby broken, i.e. right circularly polarized (RCP, +) waves and left circularly polarized (LCP, -) waves have different refractive indices. The chirality parameter $\kappa$ describes the strength of the cross-coupling effect, so that the constitutive relations of a chiral medium is given by

$$\begin{pmatrix} D \\ B \end{pmatrix} = \begin{pmatrix} \varepsilon_0 \varepsilon & -i\kappa/c \\ i\kappa/c & \mu_0 \mu \end{pmatrix} \begin{pmatrix} E \\ H \end{pmatrix}, \qquad (1)$$

where $\varepsilon_0$ and $\mu_0$ are the permittivity and permeability of vacuum. $\varepsilon$ and $\mu$ are the relative permittivity and permeability of the chiral medium. $c$ is the speed of light in vacuum. Assuming a time dependence of $e^{-i\omega t}$, the RCP (+) wave and LCP (-) wave are defined as $E^{\pm} = \frac{1}{2} E_0 (\hat{x} \mp i\hat{y})$ [9]. The refractive indices for RCP and LCP waves can be expressed as $n_\pm = n \pm \kappa$ [10], where $n = \sqrt{\varepsilon\mu}$. When $\kappa$ is large enough, either $n_+$ or $n_-$ becomes negative. At the same time, both RCP and LCP waves have the same impedance of $z/z_0 = \sqrt{\mu/\varepsilon}$, where $z_0$ is the impedance of the vacuum.

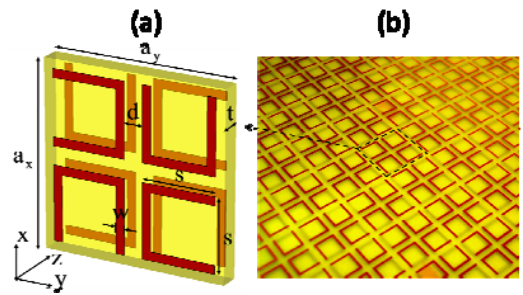

FIG. 1. (Color online) (a) Schematic of a unit cell of the chiral metamaterials consisting of four "U" split ring resonators. (b) A photo of the experimental sample. The geometric parameters are given by $a_x = a_y = 15$ mm, $t = 1.6$ mm, $d = 1.5$ mm, $w = 0.7$ mm, and $s = 6$ mm. The copper has a thickness of 0.03 mm.

Since the beginning of the proposal of metamaterials, split ring resonators (SRRs) have played an important role in achieving negative permeability. Experiments showed that SRRs can be used as magnetic metamaterials even at infrared and visible frequencies [11-13]. Furthermore, a magnetic dimer can be formed by stacking two mutually twisted SRRs, and an array of these magnetic dimers can possess optical activity [14, 15]. However, due to the lack of the rotational symmetry, the optical activity is sensitive to the linear polarization of the incident wave. To eliminate this shortage, we propose here a design in which the U-shaped SRR pairs are arranged in C4 symmetry (see Fig. 1). Thus, the constructed CMM is effectively uniaxial for the normal incidence wave. A related retrieval procedure [6,16,17] can then be applied to calculate the effective parameters for our CMMs.

Figure 1 shows the schematics of the CMMs we study. The unit cell of the CMM structure consists of double-layered four "U" split ring resonators (Four-U-SRRs) patterned on opposite sides of an FR-4 board. This structure has been mentioned theoretically [18] and is discussed in another papers [19, 20]. The relative dielectric constant of the FR-4 board is 4.0 with a dielectric loss tangent of 0.025. The two layers of Four-U-SRRs are mutually twisted by 90 degrees. The Four-U-SRRs are arranged so that the CMM structure possesses C4 symmetry. The dimensions of the unit cell and the photo of our experimental sample are shown in Fig.1 and the caption.

In order to study the behavior of the chiral structure, we conducted numerical simulations and experiments. The simulation works were carried out by using CST microwave studio (Computer Simulation Technology GmbH, Darmstadt, Germany), wherein the finite integration technique was applied. The periodic boundary conditions were applied to the $x$ and $y$ directions, and the absorbing boundary conditions were applied to the $z$ direction. In the experiment, we fabricated the chiral structures with a dimension of 20 by 20 unit cells. The transmission coefficient was measured by an HP-8510C network analyzer with two standard horn antennas.

In order to study the transmission behaviors of the chiral structure, in our simulation and experimental works, a linearly polarized electromagnetic wave (E field in the $x$ direction) is incident on the chiral structure. On the other side of the structure, we measured the transmitted field in the $x$ and $y$ polarizations ($T_{xx}$ and $T_{yx}$). Due to the four-fold rotational symmetry, circular polarization conversion is absent. The transmission of circularly polarized waves can be converted from the linear transmission coefficients $T_{xx}$ and $T_{yx}$ [6,7], $T^\pm = T_{xx} \pm iT_{yx}$. For the transmitted EM wave, the polarization azimuth rotation angle $\theta$ can be calculated as $\theta = [\arg(T^+) - \arg(T^-)]/2$. The ellipticity of the transmitted wave is defined as $\eta = \arctan[(|T^+| - |T^-|)/(|T^+| + |T^-|)]$ [9], which also measures circular dichroism.

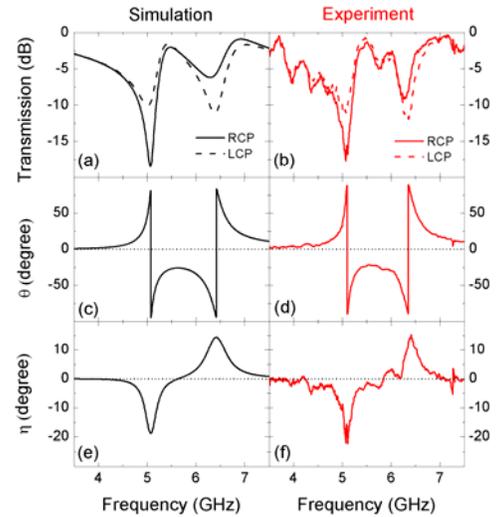

FIG. 2. (Color online) Simulation and experimental results for the chiral metamaterial. (a) and (b) are the transmission spectra for RCP and LCP waves. (c) and (d) are the polarization azimuth rotation angle $\theta$. (e) and (f) are the ellipticity angle $\eta$ of the transmitted wave.

Figure 2(a) and 2(b) show the simulated and measured transmission spectra. There are obvious differences between the transmissions of RCP and LCP waves around the resonances. Around the frequency of 5.1 GHz, the transmission of LCP wave is 7~8 dB higher than that of the RCP wave. While around the frequency of 6.3 GHz, the transmission of the LCP wave is around 3~4 dB lower than that of the RCP wave. The fluctuations on the experiment curve come from the multiple reflections between the CMM structure and the horn antennas.

Figures 2(c) to 2(f) show the simulated and experimental results of the polarization azimuth rotation angle $\theta$ and the ellipticity $\eta$ of the transmitted wave. It can be clearly seen that, in the middle of the two resonant frequencies, the ellipticity $\eta = 0$, where corresponds a pure optical activity effect, i.e. for the linear polarization incident wave, the transmission wave is still the linear polarization but with a rotated angle $\theta$. The azimuth rotation angle $\theta$ is as large as $26°$ at 5.5 GHz. The rotation angle $\theta$ per wavelength is $700°/\lambda$.

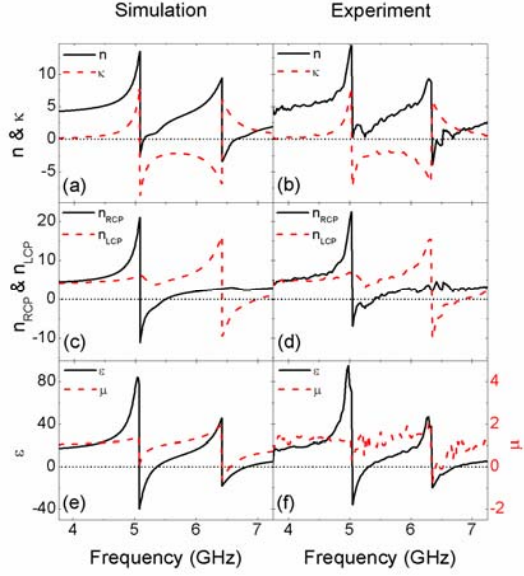

FIG. 3. (Color online) The retrieved effective parameters of the chiral metamaterials based on the simulation (left) and experimental (right) data. (a-b) show the real parts of the refractive index n and chirality $\kappa$. (c-d) shows the real parts of the refractive indices for RCP and LCP waves. (e-f) shows the real parts of the permittivity $\varepsilon$ and permeability $\mu$.

Figure 3 shows the retrieved effective parameters based on the simulation and experimental data of the transmission and reflection for one layer of the CMMs. In the retrieval process, the effective thickness of the CMM is assumed to be 1.8 mm along the wave propagation direction. The chirality is very large, e.g. it's 2.15 at $\eta = 0$. If using the following expression [20]

$$\kappa = -\frac{\Omega_{\kappa 1}\omega_{\kappa 1}\omega}{\omega^2 - \omega_{\kappa 1}^2 + i\Gamma_{\kappa 1}\omega_{\kappa 1}\omega} + \frac{\Omega_{\kappa 2}\omega_{\kappa 2}\omega}{\omega^2 - \omega_{\kappa 2}^2 + i\Gamma_{\kappa 2}\omega_{\kappa 2}\omega} \quad (2)$$

to fit the result of the chirality $\kappa$, it gives us $\Omega_{\kappa 1} = 0.22$, $\omega_{\kappa 1} = 5.08$ GHz, $\Gamma_{\kappa 1} = 0.02$, $\Omega_{\kappa 2} = 0.28$, $\omega_{\kappa 2} = 6.42$ GHz, $\Gamma_{\kappa 2} = 0.03$. Comparing Figs. 3(a-b) and Figs. 3(c-d), due to the relation of $n_\pm = n \pm \kappa$, the strong chirality $\kappa$ can push the refractive index of RCP (LCP) wave to be negative around the resonant frequency of 5.1 (6.4) GHz as shown in Fig. 3(c-d). However, due to the losses in the dielectric board, the figure of merit [–Re(n)/Im(n)] is only around 1.1 (1.3) at 5.1 (6.4) GHz in the negative index band. By using a lower loss substrate, the imaginary part of the negative index can be reduced. The figure of merit can be larger than 10 (not shown here). In Figs. 3(e-f), we show the retrieved real parts for the permittivity and permeability of the CMM. Obviously, the Re($\mu$) is always positive throughout the entire frequency range (except the tiny range around 6.4GHz), while Re($\varepsilon$) is negative in the frequency ranges above the two resonances. In traditional metamaterials, this will not result in a negative index. Therefore, the negative index of RCP (LCP) wave is actually attributed to the relatively small n and large chirality $\kappa$.

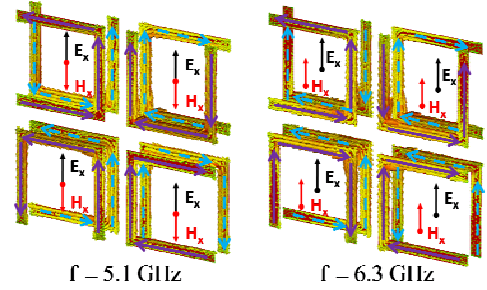

FIG. 4. (Color online) The current distributions driven by the electric field in the x direction at 5.1 GHz and 6.3 GHz. The arrows in the middle of the SRRs show the direction of the x component of the induced magnetic field.

In order to understand the mechanism of the resonances for our Four-U-SRRs CMM, we study the current modes on the Four-U-SRRs. Fig. 4 shows the simulated current modes at 5.1 GHz and 6.3 GHz driven by the electric field in the x direction. At 5.1 GHz, the currents on the top and bottom Four-U-SRRs are in the same direction. On the contrary, at 6.3 GHz, the currents on the top and bottom are in the opposite direction. According to the current distribution, we can analyze the induced magnetic field. Fig. 4 shows that the x component of the induced magnetic field $H_x$ is nonzero. And moreover, at 5.1 GHz, $H_x$ and $E_x$ are in the opposite directions, while at 6.3 GHz, they are in the same direction.

This causality between electric and magnetic fields is consistent with the constitute equation Eq. (1).

In summary, by twisting the four "U" split ring resonators in the propagation direction, which owns the fourfold rotational symmetry, a uniaxial chiral metamaterial can be constructed. We study the structure via the numerical simulation and the experiment at 3.5~7.5 GHz. The simulation results agree with the experimental results very well. This artificial structure gives us so strong chirality that the refractive index of RCP or LCP can be pushed to be negative at the vicinity of the resonance. Such structure can be scaled to other frequencies. The proposed chiral structures are suitable for planar fabrication. In the future, we can use this planar design to contracture bulk chiral metamaterials via layer-by-layer method.

This work is supported by the European Union under the projects EU-PHOME and EU-ECONAM, and TUBITAK under the Project Nos., 107A004, and 107A012. One of the authors (E.O.) also acknowledges partial support from the Turkish Academy of Sciences. Rongkuo Zhao acknowledges the China Scholarship Council (CSC) for financial support.

*References*:

1. J. B. Pendry, Science **306**, 1353 (2004).
2. S. Tretyakov, I. Nefedov, A. Sihvola, S. Maslovski, and C. Simovski, J. Electromagn. Waves Appl. **17**, 695 (2003).
3. A. V. Rogacheva, V. A. Fedotov, A. S. Schwanecke, and N. I. Zheludev, Phys. Rev. Lett. **97**, 177401 (2006).
4. B. Wang, J. Zhou, T. Koschny, and C. M. Soukoulis, Appl. Phys. Lett. **94**, 151112 (2009).
5. S. Zhang, Y.-S. Park, J. Li, X. Lu, W. Zhang, and X. Zhang, Phys. Rev. Lett. **102**, 023901 (2009).
6. E. Plum, J. Zhou, J. Dong, V. A. Fedotov, T. Koschny, C. M. Soukoulis, and N. I. Zheludev, Phys. Rev. B **79**, 035407 (2009).
7. J. Zhou, J. Dong, B. Wang, T. Koschny, M. Kafesaki, and C. M. Soukoulis, Phys. Rev. B **79**, 121104(R) (2009).
8. M. Decker, M. Ruther, C. E. Kriegler, J. Zhou, C. M. Soukoulis, S. Linden, and M. Wegener, Opt. Lett. **34**, 2501 (2009).
9. J. D. Jackson, *Classical Electrodynamics*, 3rd ed. (Wiley, New York, 1998).
10. J. A. Kong, *Electromagnetic Wave Theory* (EMW Publishing, Cambridge, MA, 2008).
11. C. Enkrich, M. Wegener, S. Linden, S. Burger, L. Zschiedrich, F. Schmidt, J. F. Zhou, Th. Koschny, and C. M. Soukoulis, Phys. Rev. Lett. **95**, 203901 (2005).
12. S. Linden, C. Enkrich, M. Wegener, J. Zhou, T. Koschny, and C. M. Soukoulis, Science **306**, 1351 (2004).
13. I. Sersic, M. Frimmer, E. Verhagen, and A. F. Koenderink, Phys. Rev. Lett. **103**, 213902 (2009).
14. H. Liu, D. A. Genov, D. M. Wu, Y. M. Liu, Z. W. Liu, C. Sun, S. N. Zhu, and X. Zhang, Phys. Rev. B. **76**, 073101 (2007).
15. N. Liu, H. Liu, S. Zhu, and H. Giessen, Nat. Photonics **3**, 157 (2009).
16. D. H. Kwon, D. H. Werner, A. V. Kildishev, and V. M. Shalaev, Opt. Express **16**, 11822 (2008).
17. C. Menzel, C. Rockstuhl, T. Paul, and F. Lederer, Appl. Phys. Lett. **93**, 233106 (2008).
18. N. Liu and H. Giessen, Opt. Express **16**, 21233 (2008).
19. X. Xiong, W. H. Sun, Y. J. Bao, M. Wang, R. W. Peng, C. Sun, X. Lu, J. Shao, Z. F. Li, and N. B. Ming, Phys. Rev. B **81**, 075119 (2010).
20. R. Zhao, Th. Koschny, E. N. Economou, C. M. Soukoulis, Phys. Rev. B **81**, 235126 (2010).